\documentclass[parskip]{scrartcl}

\usepackage[margin=15mm]{geometry}

\usepackage[margin=15mm]{geometry} \usepackage{tikz}
\tikzstyle arrowstyle=[scale=1]
\tikzstyle directed=[postaction={decorate,decoration={markings,
    mark=at position .65 with {\arrow[arrowstyle]{stealth}}}}]
\tikzstyle reverse directed=[postaction={decorate,decoration={markings,
    mark=at position .65 with {\arrowreversed[arrowstyle]{stealth};}}}]

\usepackage{tikz} \usetikzlibrary{arrows,shapes,positioning}
\usetikzlibrary{decorations.markings}
\tikzstyle arrowstyle=[scale=1.6]
\tikzstyle directed=[postaction={decorate,decoration={markings,
    mark=at position .65 with {\arrow[arrowstyle]{stealth}}}}]
\tikzstyle reverse directed=[postaction={decorate,decoration={markings,
    mark=at position .65 with {\arrowreversed[arrowstyle]{stealth};}}}]

\usepackage[hang,small,bf]{caption}

\usepackage{bm}

\usepackage{wrapfig}
\usepackage{color,epsfig,amsfonts,amssymb}

\usepackage{url}

\usepackage{amsmath}

\usepackage[pdftex,breaklinks,colorlinks,
citecolor=blue,allcolors=blue,
urlcolor=blue]{hyperref}

  \newcommand{\spce}{\;\; \: \; \; \; \; \; \;}
\newcommand{\comma}{\;\; ,}
 \newcommand{\period}{\;\; .}
 \newcommand{\negspce}{\! \! \! \! \! \! \! \! \! \! \!  }
 \newcommand{\eq}{\; = \;}
 \newcommand{\sep}{\;\; , \;\;}
  
\newcommand{\be}{\begin{equation}}
 \newcommand{\bd}{\begin{displaymath}}
 \newcommand{\ee}{\end{equation}}
 \newcommand{\ed}{\end{displaymath}}
 \newcommand{\ba}{\begin{eqnarray}}
 \newcommand{\ea}{\end{eqnarray}}
 
 \newcommand{\minus}{\! - \!}

 \renewcommand{\i}{{\mathrm i}}
 \newcommand{\e}{{\mathrm e}}

 \newcommand{\mybf}{\normalfont \bfseries}
\newcommand{\myit}{\normalfont \itshape}
\newcommand{\myrm}{\mathrm}
\newcommand{\Ut}{\tilde{U}}
\newcommand{\Vt}{\tilde{V}}
\newcommand{\Wt}{\tilde{W}}

\newcommand{\kt}{\tilde{\kappa}}
\newcommand{\nt}{\tilde{n}}
\newcommand{\Ft}{\boldsymbol{\tilde{F}}}

 \newcommand{\half}{{\scriptstyle \frac{1}{2}}}
\newcommand{\m}{\raisebox{6.2pt}{$\hspace{0.18mm} \scriptstyle m$}}
\newcommand{\md}{\raisebox{5.7pt}{$\hspace{0.18mm} \scriptstyle m$}}
\newcommand{\mdb}{\raisebox{5.7pt}{$\hspace{0.18mm} \scriptstyle 2 m$}}
\newcommand{\rs}{\raisebox{5.2pt}{$\hspace{0.18mm} \scriptstyle 2 $}}
\newcommand{\rsq}{\raisebox{5.2pt}{$\hspace{0.18mm} \scriptstyle 4 $}}
\newcommand{\rsnn}{\raisebox{5.8pt}{$\! \! \!  \scriptstyle 2 N $}}
\newcommand{\rsmm}{\raisebox{5.8pt}{$\! \! \!  \scriptstyle 2 M $}}

 \title{Bulk, surface and corner free energies of the anisotropic triangular Ising model: series expansions and critical behaviour}

 \author{ R.J. Baxter\\
 {\protect \small  Mathematical
 Sciences Institute}\\
 {\protect  \small The Australian National University,
 Canberra, A.C.T. 0200,
  Australia}}

 \date{\protect \small  23 Sep  2019}

\begin{document}


 \maketitle

 \abstract{We consider the  anisotropic Ising model on the triangular lattice with finite
 boundaries, and use Kaufman's spinor method to calculate low-temperature 
 series expansions for the 
 partition function  to high order. From these we can obtain 108-term 
 series expansions for the bulk, surface  and corner free energies. We extrapolate these
to all terms and thereby conjecture the exact results for each. 
 
 Our results agree with the exactly known bulk free energy. For the isotropic case, they 
 also agree with Vernier and Jacobsen's conjecture for the $60^{\circ}$ corners, and 
 with Cardy and Peschel's conformal invariance predictions for the dominant 
 behaviour at criticality.}






 \section{Introduction}
 Vernier and Jacobsen\cite{VJ2012}  considered a number of two-dimensional 
lattice models in statistical mechanics that are ``exactly solved" in the sense that their
bulk free energies (and where appropriate their order parameters) have been calculated
exactly. They developed series expansions of typically twenty--five  or so terms for the 
surface and corner  free energies, and from these were able to conjecture the exact forms.

 For the square lattice Ising model, the bulk free energy was obtained by Onsager in 
 1944,\cite{Onsager1944}  and the surface free energy in 1967 by McCoy and 
 Wu.\cite[eqn.4.24b]{McCoyWu67}\cite[p.126, eqn.4.24b]{MCWbook}
In 2017 the author\cite{RJB2017}   and Hucht\cite{Hucht1,Hucht2} derived the 
low-temperature form of the corner free energy and showed that Vernier and 
Jacobsen's conjectures were indeed correct for all three free energies.


Here we consider the  anisotropic ferromagnetic Ising model on the triangular lattice
and develop low-temperature ($T < T_c$) series expansions for the ordered phase. The 
bulk free energy  was obtained in 1950 Houtappel and 
others,\cite{Hout1950},\cite{Wannier1950},\cite{Husimi1950}, 
\cite{Syozi1950} and in 1964 quite elegantly by 
Stephenson.\cite{Stephenson1964} Vernier and 
Jacobsen\cite{VJ2012} were unable to obtain enough terms in their series to reliably 
conjecture  the surface and the $120^{\circ}$ corner free energies. We have used the spinor method of 
Kaufman,\cite{Kaufman1949} which greatly simplifies the problem. However,  unlike 
ref.\cite{RJB2017}, we have not solved the problem algebraically, but have obtained 
the first  108  terms in the series expansions of the surface and various corner
 free energies  of the triangular latice. When we expand the results as infinite products
in powers of the elliptic nome $p$ that naturally enters the calculation, we observe 
patterns in the exponents of period 24, and from them extrapolate the full expansion.

We believe our results  for the anisotropic surface and corner free energies to be new, 
as are those for the isotropic surface and $120^{\circ}$ corner free energies.

We find agreement with previous results, in particular the predictions of conformal invariance 
for the logarithmic divergence of the corner free energies.


\section{The Ising model}
\setcounter{equation}{0}

  Following Vernier and Jacobsen, we first consider the Ising model on the parallelogram 
  of the first figure in Fig.\ref{examples}.
  This has $M$ rows, $N$ columns and $MN$ sites (including the boundaries and corners). 
  It  also has $2N$ sites on the upper and lower horizontal boundaries, and $2M$ on the two 
  sloping boundaries. On each site $i$ we place a spin $\sigma_i$ with value $+1$ or $-1$.
  The partition function is
  \be \label{pfn}
  Z  \eq  \sum_{ \{ {\bm \sigma} \} } \exp \left[ \sum_{i,j}  K_1 \sigma_i \sigma_j  + 
  \sum_{i,k}  K_2 \sigma_i \sigma_k   +
  \sum_{i,l}  K_3 \sigma_i \sigma_l   \right]  \comma  \ee
  where the outer sum is over all $2^{MN}$ values of spins
   ${\bm \sigma} = \{ \sigma_1, \sigma_2, \ldots , \sigma_{MN} \} $, the first inner sum is 
over  all adjacent  horizontal pairs of sites (i.e. edges) $i,j$. Similarly, the second sum is 
over all  edges $i,k$ parallel to the left and right boundaries, and the third over all
 edges $i,l$ in the remaining direction. we shall refer to these three types of edges as
  types 1, 2, 3, respectively.
  
When $K_1,K_2,K_3$ are all large and positive, the largest two contributions to the 
sum in (\ref{pfn}) are from the cases when all the spins are equal, either to +1 or to -1. 
If we define $\widehat{Z}$ by 
\be \label{ZtZ}
 Z \eq 2 \, \e^{M(N-1)K_1+N(M-1)K_2 +(M-1)(N-1)K_3 }  \; {\widehat{Z} } 
\comma \ee
 then it follows that 
 \be {\widehat{Z}} \eq 1 + {\mathrm {smaller \; \; terms } } \comma \ee
 where for given $M,N$  the smaller terms tend to zero as   $K_1,K_2,K_3
 \rightarrow \infty$.

Considering the effect of changing the sign of just a few of the spins, defining 
the three Boltzmann weights
\be \label{defzj}
z_j = e^{-2 K_j} \; \; \; \; {\mathrm{for} \; \; } j=1,2,3 \; , \ee
and expanding in the combined powers of $z_1, z_2, z_3$,  we find that
\ba {\widehat{Z}}  && \negspce =    1+2 z_1z_2 +2 z_1 z_2 z_3  + 3 z_1^2 z_2^2   +
(2M-2) z_1 z_2^2 z_3 +(2N-2) z_1^2 z_2 z_3 +  \nonumber \\
&& 4 z_1^2 z_2^2 z_3 + 2 z_1^2 z_2 z_3^2  + 2 z_1 z_2^2 z_3^2 + 6 z_1^3 z_2^3 
+ (4 M - 4) z_1^2 z_2^3 z_3 + \\ 
&& (4 N - 4) z_1^3  z_2^2  z_3 +   2 z_1^3 z_2 z_3^2 + (M N - 5) z_1^2 z_2^2 z_3^2 
+  2 z_1 z_2^3 z_3^2   + \cdots  \nonumber  \ea
so for the isotropic case $z_1 = z_2 = z_3  = z$:
\bd {\widehat{Z}}  =    1+2 z^2+2 z^3 + (2M+2N-1)z^4 + 8z^5 + (MN+4M+4N-3)z^6 + 
\mathrm{O} (z^7) \ed

\noindent These results  are correct to this order $z^6$ for $M,N  \geq  4  $. To order $z^4$ 
they are correct for $M, N  \geq 3 $. 
     

     Developing these series further, and expanding $\log  {\widehat{Z}} $  rather than 
${\widehat{Z}} $, we find that to all the orders we have calculated,   $\log  {\widehat{Z}} $ is
{\em{ linear}}  in $M$ and $N$, provided $M, N$  are sufficiently large. This implies that 
\be  \label{specialZ}  {\widehat{Z}}  =  \kappa_b^{MN}  \,  \kappa_{s,1}{\rsnn} \,   
\kappa_{s,2}{\rsmm}\, \kappa_c^4 \comma \ee
where $\kappa_b, \kappa_{s_1},  \kappa_{s_2},  \kappa_c$ are {\em independent} of $M,N$.
These are the bulk, surface and corner free energies partition functions  discussed in 
\cite{RJB2017} . They are the exponential of the free energies discussed by 
Vernier and Jacobsen.\cite{VJ2012} 


\setlength{\unitlength}{1pt}
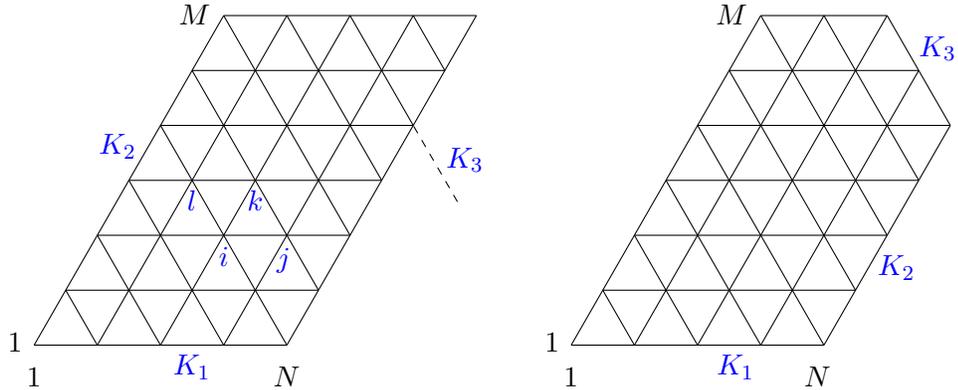
\begin{figure}[hbt]
\begin{picture}(420,240) (-60,-5)

\begin{tikzpicture}[scale=0.84]
\draw[thin] (0,0) -- (4,0);
\draw[thin] (0.5,0.87) -- (4.5,0.87);
\draw[thin] (1,1.74) -- (5,1.74);
\draw[thin] (1.5,2.61) -- (5.5,2.61);
\draw[thin] (2,3.48) -- (6,3.48);
\draw[thin] (2.5,4.35) -- (6.5,4.35);
\draw[thin] (3,5.22) -- (7,5.22);

\draw[thin] (1,0) -- (0.5,0.87);
\draw[thin] (2,0) -- (1,1.74);
\draw[thin] (3,0) -- (1.5,2.61);
\draw[thin] (4,0) -- (2,3.48);
\draw[thin] (4.5,0.87) -- (2.5,4.35);
\draw[thin] (5,1.74) -- (3,5.22);
\draw[thin] (5.5,2.61) -- (4,5.22);
\draw[thin] (6,3.48) -- (5,5.22);
\draw[thin] (6.5,4.35) -- (6,5.22);

\draw[thin] (0,0) -- (3,5.22);
\draw[thin] (1,0) -- (4,5.22);
\draw[thin] (2,0) -- (5,5.22);
\draw[thin] (3,0) -- (6,5.22);
\draw[thin] (4,0) -- (7,5.22);

\draw[dashed,thin] (6.7,2.27) -- (6,3.48);

\node at (0,-0.5) {$1$};
\node at (4,-0.5) {$N$};

\node at (-0.3,0.05) {$1$};
\node at (2.54,5.24) {$M$};

 {\color{blue} 
  \node at (3.95,1.35) {$j$};
 \node at (3.0,1.4) {$i$};
  \node at (3.5,2.3) {$k$};
   \node at (2.5,2.3) {$l$};

\node at (2.5,-0.35) {$K_1$};
\node at (1.32,3.15) {$K_2$};
\node at (6.82,2.94) {$K_3$};

\node at (11.1,-0.35) {$K_1$};
\node at (13.65,1.23) {$K_2$};
\node at (14.3,4.7) {$K_3$};

}

\node at (8.5,-0.5) {$1$};
\node at (12.4,-0.5) {$N$};

\node at (8.2,0.05) {$1$};
\node at (11.04,5.24) {$M$};

\draw[thin] (8.5,0) -- (12.5,0);
\draw[thin] (9,0.87) -- (13,0.87);
\draw[thin] (9.5,1.74) -- (13.5,1.74);
\draw[thin] (10,2.61) -- (14,2.61);
\draw[thin] (10.5,3.48) -- (14.5,3.48);
\draw[thin] (11,4.35) -- (14,4.35);
\draw[thin] (11.5,5.22) -- (13.5,5.22);

\draw[thin] (8.5,0) -- (11.5,5.22);
\draw[thin] (9.5,0) -- (12.5,5.22);
\draw[thin] (10.5,0) -- (13.5,5.22);
\draw[thin] (11.5,0) -- (14,4.35);
\draw[thin] (12.5,0) -- (14.5,3.48);

\draw[thin] (14.5,3.48) -- (13.5,5.22);
\draw[thin] (14,2.61) -- (12.5,5.22);
\draw[thin] (13.5,1.74) -- (11.5,5.22);
\draw[thin] (13,0.87) -- (11,4.35);
\draw[thin] (12.5,0) -- (10.5,3.48);
\draw[thin] (11.5,0) -- (10,2.61);
\draw[thin] (10.5,0) -- (9.5,1.74);
\draw[thin] (9.5,0) -- (9,0.87);

 \end{tikzpicture}
 \end{picture}

 \vspace{0.5cm}

  \caption{ Examples of quadrilateral shapes on the the triangular lattice}

 \label{examples}
\end{figure}


\section{Transfer matrices}
\setcounter{equation}{0}

{From} now on, we consider $\widehat{Z}$ rather than $Z$, i.e. we remove the leading factor
$\e^{K_i}$ in each edge of type $i$.
    
   We can construct the partition function of the first lattice in Fig.\ref{examples} in the 
usual way, using  transfer matrices. We follow the notation of section 2 of \cite{RJB2017}, 
except that here we use local transfer matrices that each add only a single edge of the 
lattice, and we re-arrange the $2N$-dimensional matrices so that rows $1,2,3, \ldots , N$ 
become  $1,3, 5 ,\ldots , 2N-1$ and  rows $N+1,N+2, \ldots , 2N$ become  
$2,4, \ldots , 2N$, and similarly for the columns.


   Let $\sigma = \{ \sigma_1, \ldots ,\sigma_N \}$ be the spins on a row of the lattice,
and  $\sigma' = \{ \sigma'_1, \ldots ,\sigma'_N \}$ be the spins on the row above. Then
we  can define the $2^N$-dimensional row-to-row transfer matrices $U_{j,j+1},V_j, W_{j,j+1}$, 
with elements
\bd
\left( U_{j,j+1} \right)_{\sigma,\sigma'} \eq  \e^{K_1( \sigma_j \sigma_{j+1}  -1)} \, 
\delta_{ {\bm \sigma}  , {\bm \sigma'}  }  \ed
\be \label{defUVW}
\left( V_{j } \right)_{\sigma,\sigma'} \eq  \left[ \delta_{\sigma_j ,\sigma'_j} + 
 \e^{-2 K_2} \delta_{\sigma_j ,-\sigma'_j} \right]
\, \prod_{k=1, k \neq  j}^N \delta_{ \sigma_k  , \sigma'_k}   \ee
\bd \left( W_{j,j+1} \right)_{\sigma,\sigma'} \eq  \e^{K_3 (\sigma_j \sigma_{j+1} -1)} \, 
\delta_{ {\bm \sigma}  , {\bm \sigma'}  }  \ed
 The matrices $U_{j,j+1}, W_{j,j+1}$ are diagonal.
 

 Let 
\be {\mathbf 1}  \eq  \left(  \begin{array}{cc}
1 & 0 \\
0 & 1 \end{array}  \right) \sep
{\mathbf s} \eq  \left(  \begin{array}{cc}
1 & 0 \\
0 & -1 \end{array}  \right) \sep
{\mathbf c} \eq  \left(  \begin{array}{cc}
0 & 1 \\
1 & 0 \end{array}  \right) \comma \ee
and define $s_j, c_j$ to be the $2^N$-dimensional matrices
\ba s_j \eq  {\mathbf 1}  \otimes  \cdots \otimes {\mathbf 1} \otimes {\mathbf s}
 \otimes {\mathbf 1} \otimes 
\cdots  \otimes {\mathbf 1} \comma \nonumber \\
c_j \eq  {\mathbf 1}  \otimes  \cdots \otimes {\mathbf 1} \otimes {\mathbf c} 
\otimes {\mathbf 1} \otimes 
\cdots  \otimes {\mathbf 1} 
\comma \ea
$\mathbf{s}$, $\mathbf{c}$  on the RHS being in position $j$. Then
\be  \negspce \negspce U_{j,j+1} \eq \exp \{ K_1 (s_j s_{j+1} -1) \}  \comma \ee
\be   \label{defV1}
V_j \eq (1- \e^{-4 K_2} )^{1/2} \; \e^{ K^*_2 c_j    }  \ee
\be  \negspce \negspce  W_{j,j+1}  \eq \exp \{ K_3 (s_j s_{j+1}  -1) \}  \comma \ee
where 
\be \label{defH*}
\tanh K^*_2 \eq \e^{-2K_2} \period \ee

The usual row-to-row transfer matrix is
\be  \label{defT}
T \eq U_{1,2} V_1 W_{1,2} \cdots U_{N-1,N} V_{N-1}  
W_{N-1,N} \, V_N  \ee
(each $U$ adds a horizontal edge in the lower row, each $V$ a vertical edge , and 
each $W$ a slanting edge). Let $\xi$ be the $2^N$-dimensional vector with all entries 
unity. Then the partition function is 
\be Z \eq \xi^T  T^M  U_1 U_2 \cdots U_N \xi \ee
(here the $U's$ add the bottom row of the lattice).


As in Kaufman\cite{Kaufman1949} and \cite{RJB2017}, we can replace the 
$2^N$-dimensional matrices  $U, V, W$ by $2N$-dimensional ones
 $\Ut, \Vt, \Wt$, with elements
\be (\Ut_{m,m+1} )_{i,j} = (\Vt_{m} )_{i,j} = (\Wt_{m,m+1} )_{i,j}  =  
 \delta_{i,j} \; \; \mathrm{for} \; \;  i,j = 1, \ldots , 2N \comma  \ee
except for the entries
\bd  (\Ut_{m,m+1} )_{2m,2m} =  (\Ut_{m,m+1} )_{2m+1,2m+1} = \cosh 2 K_1
  \comma \ed
\bd  (\Ut_{m,m+1} )_{2m,2m+1} =  - (\Ut_{m,m+1} )_{2m+1,2m} = \i \sinh 2 K_1  
\comma \ed
\bd  (\Vt_{m} )_{2m-1,2m-1} =  (\Vt_{m} )_{2m,2m} = \cosh 2 K_2/ \sinh 2 K_2  
 \comma \ed
\bd  (\Vt_{m} )_{2m-1,2m} =  - (\Vt_{m} )_{2m,2m-1} = -\i /\sinh 2 K_2  \comma \ed
\bd  (\Wt_{m,m+1} )_{2m,2m} =  (\Wt_{m,m+1} )_{2m+1,2m+1} = \cosh 2 K_3 
 \comma \ed
\bd  (\Wt_{m,m+1} )_{2m,2m+1} =  - (\Wt_{m,m+1} )_{2m+1,2m} = \i \sinh 2 K_3  
\period \ed
Thus each of the $\Ut, \Vt, \Wt$ matrices has just  four elements that are not 1 or 0.

If we also define the simple $2N$ by $N$ matrix  $J$ with entries zero except for 
\bd J_{2j-1,j} = 1 \sep  J_{2j,j} = \i  \; \; \mathrm{for} \; \; j=1, \ldots , N  \comma \ed
and, analogously to (\ref{defT}),
\be \label{deftT} \tilde{T}  \eq \Ut_{1,2}  \,  \Vt_1 \Wt_{1,2} \cdots \Ut_{N-1,N} \Vt_{N-1} \,
\Wt_{N-1,N}  \Vt_{N} 
\comma  \ee
then
\be  \label{formZ}  \widehat{Z } \eq 2^{N/2} \, \e^{-{\cal N}_1 K_1 -{\cal N}_3 K_3} \, 
 ( 1-\e^{-4K_2} )^{{\cal N}_2/2} \; (\det Q )^{1/2} 
\comma  \ee
where ${{\cal N}_2} $ is the number of  type-2 edges (in this case ${\cal N}_2 = N(M-1)$), 
and $Q$ is the $N$ by $N$ matrix 
\be \label{formula} Q \eq J^\dagger \,  \tilde{T}^M \,   \Ut_{1,2} \Ut_{2,3} \cdots 
\Ut_{N-1,N} \,  J   \comma \ee
$J^{\dagger}$ being the hermitian transpose of $J$.

\subsection*{Other shapes} 
Each of the matrices $ \Ut_{m,m+1},  \Vt_{m} ,  \Wt_{m,m+1}$ entering (\ref{formula}) via
(\ref{deftT}) can be associated with a particular edge of the first graph in 
Fig. {\ref{examples}}. Other shapes can be obtained 
by simply deleting the corresponding matrices. For instance the second graph in 
Fig. {\ref{examples}} can be obtained by deleting the  $ \Ut_{N-2,N-1},  \Ut_{N-1,N}$ 
before the $J$ in   (\ref{formula}), the $\Ut_{N-1,N} , \Vt_{N-1},  \,\Wt_{N-1,N},  \Vt_{N} $
 in the furthest-right  matrix $ \tilde{T} $ in  (\ref{formula}), and the $\Vt_{N} $ in 
 the next-furthest-right $ \tilde{T} $.

This reduces the number of  type-1 edges by 3, of  type-2 edges by 3, and of type-3 
edges  by 1, so we must  replace ${\cal N}_2$ in (\ref{formZ}) by $N(M-1)-3$.





\section{The isotropic case}
\setcounter{equation}{0}
We first focus the isotropic case, when $K_1 = K_2 = K_3 $ and $z_1 = z_2 = z_3 = z$.

We have performed  calculations on various convex shaped graphs on the triangular 
lattice,  expanding $\widehat{Z}$ to orders as high as 108 in $z$ (or in the elliiptic
 parameter  $p$
defined below). In all cases we find that (\ref{specialZ}) generalizes to
\be \label{defkappas}
\widehat{Z} \eq  2 \,  (\kappa_b)^{n_{b}} \, (\kappa_{s})^{n_s}  \,(\kappa_{c})^{n_c} \, 
(\kappa_{c})^{\nt_c} \ee
 where ${n_{b}} $ is the number of sites in the graph, $n_s$ is the number of
surface sites,  $n_c$ is the number of 
 $60^{\circ}$ corners, and  $\nt_c$ is the number of $120^{\circ}$   corners.
 
 When counting these numbers, $n_{b}$ includes {\em all} sites, including those on the 
 surfaces and corners, and  $n_s $ is the sum of {\em all} surface sites, 
 including  the adjacent corners. (So any individual surface must have at least two sites.) 
 Hence the number of distinct boundary sites on a graph is
 $n_s-n_c-\nt_c $.
 
 So for the particular graphs in Fig.\ref{examples}  (both with $M= 7, N=5$), for the first 
 graph
 \bd \{ n_b, n_s, n_c, \nt_c \} = \{ 35,24,2,2 \}   \ed
 while for the second graph 
  \bd \{ n_b, n_s, n_c, \nt_c \} = \{ 32,23,1,4 \} \ed
  and for the two graphs in Fig.\ref{trihex},
   \bd \{ n_b, n_s, n_c, \nt_c \} = \{ 28,21,3,0 \}    \; \; {\myrm{ and}}
     \; \; \{ 37,24,0,6 \}  \ed
     

 We expect
(\ref{defkappas}) to be true  in the sense that the ratio of the 
LHS to the RHS $\rightarrow 1$ when the area and all surfaces become large. Then this limit 
defines $ \kappa_b,  \kappa_{s} , \kappa_{c}, \kt_{c}$.
 

\setlength{\unitlength}{1pt}
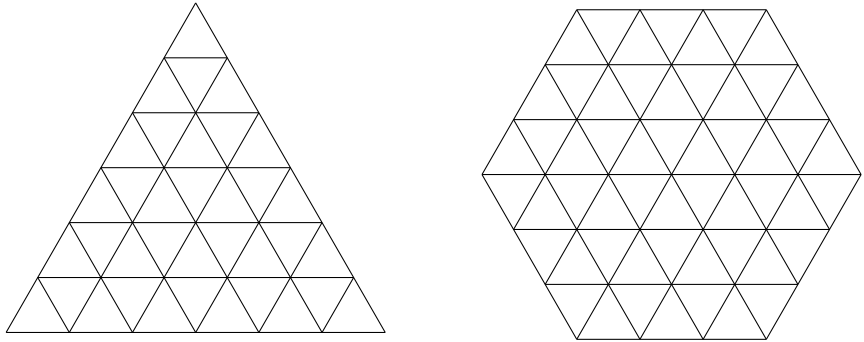
\begin{figure}[hbt]
\begin{picture}(420,240) (-90,-5)

\begin{tikzpicture}[scale=0.84]
\draw[thin] (0,0) -- (6,0);
\draw[thin] (0.5,0.87) -- (5.5,0.87);
\draw[thin] (1,1.74) -- (5,1.74);
\draw[thin] (1.5,2.61) -- (4.5,2.61);
\draw[thin] (2,3.48) -- (4,3.48);
\draw[thin] (2.5,4.35) -- (3.5,4.35);


\draw[thin] (1,0) -- (0.5,0.87);
\draw[thin] (2,0) -- (1,1.74);
\draw[thin] (3,0) -- (1.5,2.61);
\draw[thin] (4,0) -- (2,3.48);
\draw[thin] (5,0) -- (2.5,4.35);
\draw[thin] (6,0) -- (3,5.22);


\draw[thin] (0,0) -- (3,5.22);
\draw[thin] (1,0) -- (3.5,4.35);
\draw[thin] (2,0) -- (4,3.48);
\draw[thin] (3,0) -- (4.5,2.61);
\draw[thin] (4,0) -- (5,1.74);
\draw[thin] (5,0) -- (5.5,0.87);

 \end{tikzpicture}
 \end{picture}
 
 \begin{picture}(420,-90) (-270,-16)

\begin{tikzpicture}[scale=0.84]
\draw[thin] (3,0) -- (6,0);
\draw[thin] (2.5,0.87) -- (6.5,0.87);
\draw[thin] (2,1.74) -- (7,1.74);
\draw[thin] (1.5,2.61) -- (7.5,2.61);
\draw[thin] (2,3.48) -- (7,3.48);
\draw[thin] (2.5,4.35) -- (6.5,4.35);
\draw[thin] (3,5.22) -- (6,5.22);

\draw[thin] (3,0) -- (1.5,2.61);
\draw[thin] (4,0) -- (2,3.48);
\draw[thin] (5,0) -- (2.5,4.35);
\draw[thin] (6,0) -- (3,5.22);
\draw[thin] (6.5,0.87) -- (4,5.22);
\draw[thin] (7,1.74) -- (5,5.22);
\draw[thin] (7.5,2.61) -- (6,5.22);

\draw[thin] (1.5,2.61) -- (3,5.22);
\draw[thin] (2,1.74) -- (4,5.22);
\draw[thin] (2.5,0.87) -- (5,5.22);
\draw[thin] (3,0) -- (6,5.22);
\draw[thin] (4,0) -- (6.5,4.35);
\draw[thin] (5,0) -- (7,3.48);
\draw[thin] (6,0) -- (7.5,2.61);
  
 \end{tikzpicture}
 \end{picture}
 \vspace{0.5cm}

  \caption{ The regular triangular  and hexagonal  shapes on the triangular lattice}

 \label{trihex}
\end{figure}

It is convenient to expand, not in powers of the low-temperature variable $z$, but in the 
related variable $p$, defined by
\be \label{defz}
z \eq p \, \prod_{k=1}^{\infty} \frac{(1-p^{24 k-20} ) (1-p^{24 k-4})}{(1-p^{24 k-16} )
 (1-p^{24 k-8})} \period \ee

Then from (11.7.14) of \cite{mybook}, the bulk free energy is
\ba \label{kpb} \kappa_b \ & = &  \prod_{k=1}^{\infty} \frac{(1-p^{24 k-12})^2}
{(1-p^{24 k-18})(1-p^{24 k-6})} 
\times \prod_{k=1}^{\infty} \left \{  \frac{(1-p^{24 k-14})}{(1-p^{24 k-10})} \right \} ^{6 k-3} 
\nonumber \\
&& \times  \prod_{k=1}^{\infty}  \left\{ \frac{(1-p^{24 k+8}) (1-p^{24 k-2}) (1-p^{24 k-4})}
{(1-p^{24 k-8})(1-p^{24 k+2}) (1-p^{24 k+4})} \right \} ^{6k}  \comma \ea
which is the result (56) of \cite{VJ2012}.{\footnote{The $q$ of Vernier and Jacobsen
 is our $p^{3}$. We have corrected the exponent in the second product of their 
 formula  to $6k-3$.}}
 



 Presumably  the surface free energy could be calculated from the 
eigenvalues of the row-to-row transfer matrix, as has been done for the square 
lattice.\cite{McCoyWu67},\cite{MCWbook},\cite{RJB2017}  We shall not attempt this here, 
but merely present our results obtained by extrapolation from series expansions.
 
The exponents in (\ref{kpb}) are either constants or linear in $k$, and have a repeat 
pattern of period 24. For both the surface and free energies of the parallelogram in the 
first diagram of Fig. \ref{examples}, we observe that the corresponding exponents in the 
product expansions behave similarly (to the order $p^{108}$ that we calculated), except 
that some of the corner free energy exponents are quadratic in $k$. These observations 
enable us to confidently conjecture that
\ba \label{kps} 
 \negspce && \kappa_{s_1}   =  \kappa_{s_2}   = \kappa_{s_3}   = \prod_{k=1}^{\infty}  
 \left\{ \frac{1-p^{12 k-2}}{1-p^{12 k+2}} \right\}^{2k} 
 \left\{ \frac{1-p^{12 k-4}}{1-p^{12 k-8}} \right\}^{2k-1} \nonumber \\
 &  \times &  \prod_{k=1}^{\infty}  \left\{ \frac{(1-p^{24 k-2})(1-p^{24 k+10}}
 {(1-p^{24 k+2})(1-p^{24 k-10}) } \right\}^{k}
  \left\{ \frac{1-p^{24 k-16}}{1-p^{24 k-8}} \right\}^{2k-1} \\
 \times && \negspce \prod_{k=1}^{\infty} 
  \left\{ \frac{(1-p^{24 k+2})(1-p^{24 k+4})(1-p^{24 k-8})}
{(1-p^{24 k-2})(1-p^{24 k-4})(1-p^{24 k+8})} \right\}^{2k}
\left\{ \frac{1-p^{24 k-10}}
{1-p^{24 k-14}} \right\}^{2k-1} \nonumber \ea



\bd  \kappa_{c} =   \prod_{k=1}^{\infty}   \frac{(1-p^{12k-2})^{2k-1} (1-p^{24 k -16})^{5k-3} \,
 (1-p^{24 k -4})^{3k} }
{(1-p^{24k-12})^{1/3} (1-p^{12k-10}) ^{2k-1} (1-p^{24 k -20})^{3k-3} \, (1-p^{24 k -8})^{5k-2}} 
 \nonumber 
  \period \ed


\ba  {\kt_c}  & = &  \prod_{k=1}^{\infty}   \frac{(1-p^{24k-14})^{1/2} \, 
(1-p^{24k-10})^{1/2} }{(1-p^{24k-12})^{1/6}  }
\nonumber \\
& \times & \frac{1}{(1-p^{12k-9})  \, (1-p^{12k-7})^2 \, (1-p^{12k-5})^2 \,  (1-p^{12k-3}) } 
\nonumber \\
& \times & \prod_{k=1}^{\infty}  \frac{(1-p^{24 k -20})^{k^2-k+2} \, (1-p^{24 k -12})^{k^2-k-1} 
 \, (1-p^{24 k -4})^{k^2-k+2}}
{(1-p^{24 k -16})^{k^2-1} \, (1-p^{24 k -8})^{k^2-2k} \, (1-p^{24 k})^{k^2}}  \nonumber \\
& \times & \prod_{k=1}^{\infty}  
\frac{(1-p^{24 k -18})^{k+2} \, (1-p^{24 k -10})^{2k}}{(1-p^{24 k -14})^{2k-2} \, 
(1-p^{24 k -6})^{k-3}}
  \period \ea 
  


These results are true to order $p^{108}$ for all the four shapes listed here, as well as some 
others. Our formula for $\kappa_{c}$  agrees with the conjecture (60) of \cite{VJ2012}.


\subsection*{Summation formulae}

The above products can readily be converted to exponentials of sums by using the general
 formula, true for all $r,s, a,b,c $ ($r$ positive):



\ba  &&  \log  \; \prod_{k=1}^{\infty} (1-p^{rk-s})^{a+bk+ck^2} \eq   \nonumber \\
&& -  \sum_{m=1}^{\infty}  \frac{p^{(r-s)m} }{m} \,
\left[ \frac{a}{1-p^{rm}} +\frac{b-c}{(1-p^{rm})^2}+ \frac{2 c}{(1-p^{rm})^3} \right] \period \ea

Thus (\ref{kpb}) becomes
\be  \label{logk}
\log \kappa_b \eq \sum_{m=1}^{\infty} \frac{ F(p^m) }{ m  } \comma \; 
\myrm{where}  \; \; 
 F(p)  = \frac{p^6 (1-p^2)^3 (1+p^2)}{(1+p^{12})(1-p^2+p^4)^2} \period \ee

Similarly, the logarithms of $\kappa_s, \kappa_{c}, \kt_c$ are given by 
 (\ref{logk}),  with the function $F$ replaced by

\be \log \kappa_s :  \; \; F(p)  \eq \frac{p^4 (1-p^2)^2 (1-p^4) 
 (1+p^6+p^{12})}{(1+p^{12})(1-p^2+p^4)^2 (1-p^4+p^8) } \comma \ee

 \setlength{\jot}{4mm}

 \ba && \negspce \negspce  \log \kappa_{c} :  \; \;  F(p)  \eq  \frac{p^{12}}{3 (1-p^{24})} +  
  \frac{p^2 (1+p^4)(1+p^{12}) }{ (1+p^4+p^8)  (1-p^{12}) } \; \;  -  \nonumber  \\
&& \spce \spce \spce 
 \frac{p^8 (2-p^8 +3 p^{12}-p^{16}+2 p^{24})}{ (1+p^8+p^{16})  (1-p^{24}) } \comma \ea
 



 \ba \label{fkct}
 && \log \kt_c:  \; \;  F(p)  \eq  \frac{p^{12}}{6 (1-p^{24})} -  
  \frac{p^{10}}{2 (1-p^4)(1+p^8+p^{16})} +   \\
&& \frac{p^3}{(1-p^2)(1-p^2+p^4)} -  \frac{p^6(3+6 p^4+8 p^8+9 p^{12} +
8 p^{16} + 6 p^{20} +3 p^{24})}{ (1+p^4)(1+p^4+p^8)(1-p^{24}) } -\nonumber \\
&& \frac{p^4 (2+4 p^4 +3 p^8 +3 p^{12}+6 p^{16}+7  p^{20}+6 p^{24} +  3 p^{28} + 
3 p^{32} +4 p^{36}+2 p^{40})}{ (1+p^4)^2 (1+p^8+p^{16})  (1-p^{24}) }   \nonumber \ea



In every case, $F(p^{-1} ) = -F(p) $, which means that the sum over $m$  in  (\ref{logk}) 
can be extended  from $-\infty$ to $\infty$ (excluding zero), which is a valuable check 
on our conjectures  and makes it simple to use a Poisson 
transformation to expand about the critical point $p = 1$.



\section{The general anisotropic case}
\setcounter{equation}{0}

There are now three types of  surfaces: we can say that those lying on edges with
 interaction  coefficients $K_i$ are of ``type $i$" and associate with them a surface
free energy $\kappa_{s,i}$. 

There are also  three types of $60^{\circ}$  corners. If the adjoining boundary 
edges are of 
types $i-1$ and $i+1$ (mod 3), we also say they are of ``type $i$" and associate  a
 corner free energy $\kappa_{c,i} $.  Similarly, for $120^{\circ}$ corners, we 
 associate  a  corner free energy $\kt_{c,i} $. 
 
If there are $n_{s,i}, n_{c,i}, \nt_{c,i} $ surfaces, $60^{\circ}$ corners  and 
$120^{\circ}$ corners of type $i$, repectively, then
(\ref{defkappas}) generalizes to
 
 \be \label{defkappas1}
\widehat{Z} \eq  2 \,  (\kappa_b)^{n_{b}} \, \prod_{i=1}^3 (\kappa_{s,i})^{n_{s,i}} 
(\kappa_{c,i})^{n_{c,i} }  \, (\kt_{c,i})^{\nt_{c,i}} \ee

These $\kappa_b, \kappa_{s,i}, \kappa_{c,i}, \kt_{c,i}$ are the partition functions per site,
 edge-site, $60^{\circ}$-corner and $120^{\circ}$ corner, respectively. We shall 
 occasionally also use the corresponding free 
 energies  $f_b, f_{s,i}, f_{c,i}, \tilde{f}_{c,i}$, defined by
 \bd -\beta f_b = \log \kappa_b \sep  -\beta f_{s,i} = \log \kappa_{s,i} \sep  
  -\beta f_{c,i} = \log \kappa_{c,i}  \sep  -\beta \tilde{f}_{s,i} = \log \kt_{s,i} \comma \ed
  where $\beta = 1/(k_B T) $, $k_B$ being Boltzmann's constant and $T$ the temperature.

Again,  $n_{s,i} $ is the sum of {\em all} sites on surfaces of type $i$, 
 {\em including } the adjacent corners. Hence the number of distinct boundary sites on a 
 graph is  $\sum_{i=1}^3 (n_{s,i} -n_{c,i} -\nt_{c,i}) $.
 
 So for the first graph in Fig.\ref{examples}  (with $M= 7, N=5$), 
  \bd \{ n_b; n_{s,1}, n_{s,2}, n_{s,3}; n_{c,1},n_{c,2}, n_{c,3}; \nt_{c,1} ,
   \nt_{c,2} , \nt_{c,3} \} = \{ 35;10,14,0;0,0,2;0,0,2 \}   \ed
and for  the second graph in Fig.\ref{examples}   and the two graphs in 
Fig.\ref{trihex},  the corresponding numbers are
\bd  \{ 32;8,12,3;0,0,1;1,1,2 \} \, , \; \;  \{ 28;7,7,7;1,1,1;0,0,0 \}  \, , \; \; 
 \{ 37;8,8,8;0,0,0;2,2,2 \} \ed
respectively.
     

We do now encounter a problem with defining the corner free energies. We can vary $n_b$ 
and the $n_{s,i}$ arbitrarily, so (\ref{defkappas1}) does indeed define $\kappa_b$ and the 
$\kappa_{s,i}$. However, we are not free to choose any values of  $n_{c,i}$ and $\nt_{c,i}$: 
there are only 11 distinct possible convex shapes and they all satisfy
\bd n_1-\nt_1 \eq n_2 - \nt_2 \eq n_3 -  \nt_3   \period \ed
It follows that we can multiply each $\kappa_{c,i}$ by any function
$\rho_i$, provided we also divide each $\kt_{c,i}$ by $\rho_i$, and that 
$ \rho_1  \, \rho_2  \, \rho_3 \eq 1 $, i.e we are free to make the transformation
\be \label{transfm}
\kappa_{c,i} \rightarrow \rho_i  \, \kappa_{c,i} \sep
\kt_{c,i} \rightarrow \kt_{c,i} /\rho_i \sep \rho_1  \, \rho_2  \, \rho_3 \eq 1  \period \ee

Put another way, the $\kappa_b,   \kappa_{s,i},  \kappa_{c,i},  \kt_{c,i}$
only enter (\ref{defkappas1}) via the combined factors
\be \label{kapsums}
\kappa_b, \; \; \kappa_{s,i}, \; \;  \kappa_{c,i} \,\kt_{c,i} \,  , \; \;
 \kappa_{c,1}\, \kappa_{c,2} \,  \kappa_{c,3} \,  , \; \;  \kt_{c,1}\, \kt_{c,2} \,  \kt_{c,3} \,  ,  \ee
for $ i = 1, 2, 3$.

We have made a number of calculations for various convex shapes (up to 108 terms in 
expansions in powers of $p$). They all support the conjecture that (\ref{defkappas1}) is 
true in the limit when each surface is large,  the $ \kappa_b, 
  \kappa_{s,i} , \kappa_{c,i},  \kt_{c,i}$ being {\em independent} 
of the shape.

\subsection*{Elliptic function parametrization}

To handle the anisotropic case, it is convenient to introduce two ``elliptic-type'' functions
 \be \label{defGwq}
 G(w,q) = \prod_{n=1}^{\infty} \frac{(1-q^{4n-3} /w )(1-q^{4n-1} w)}{(1-q^{4n-3} w )
(1-q^{4n-1}/ w)}\comma \ee
\be \label{defHwq} H(w,q) =  \prod_{n=1}^{\infty} \left[ \frac{(1-q^{2n-1} /w )}{(1-q^{2n-1} w )} \right]^{2n-1}
 \left[ \frac{(1-q^{2n} w)}{(1-q^{2n}/ w)} \right]^{2n}\comma \ee
so
\be \label{logG} 
 \log \, G(w,q) \eq \sum_{m=1}^{\infty} \frac{q^m(w^m- w^{-m} )}{m(1+q^{2 m})}\comma \ee
\be \label{logH}
\log \, H(w,q) \eq \sum_{m=1}^{\infty} \frac{q^m(w^m- w^{-m} )}{m(1+q^{m})^2}
\period \ee

We also introduce parameters $q, a_1,a_2,a_3$
such that
\be \label{aprod}
a_1 \, a_2 \, a_3 = q \ee
and the $z_1, z_2, z_3$ of (\ref{defzj}) are
\be \label{defz2}
z_j \eq z(a_j,q) \eq a_j^{1/2} \, G(a_j,q) \period \ee

For the ferromagnetic case, we must have $ 0 < z_j \leq  1$, so
each of $a_1, a_2, a_3 $ must be less than one. Then (\ref{aprod}) implies
that for $j =1, 2, 3$,
\be \label{restra}
q < a_j  \leq   1 \period \ee

The anisotropic triangular Ising model is discussed in section 11.7 of 
\cite{mybook}. Here we write the $l, K'_1,K'_2,K'_3$ therein as $k, K_1,K_2,K_3$.
Then from  eq. (11.7.5) therein, with $v_j' = (1-z_j)/(1+z_j)$, we obtain
\bd  k^2 \eq \frac{16 z_1^2 z_2^2 z_3^2 }{(1+z_2 z_3+z_3 z_1+z_1 z_2)
(1+z_2 z_3-z_3 z_1-z_1 z_2)} \;  \times \ed
\be  \label{ksq}  \frac{1}{(1-z_2 z_3+z_3 z_1-z_1 z_2)
(1-z_2 z_3-z_3 z_1+z_1 z_2))} \period \ee

This $k$ is the elliptic modulus corresponding to the nome $q$. From (15.1.4) of 
\cite{mybook}, 
\be k \eq 4 q^{1/2} \prod_{n=1}^{\infty} \left( \frac{1+q^{2n}}{1+q^{2n-1} } \right)^4 
\eq  \prod_{n=1}^{\infty} \left( \frac{1-{q'}^{2n-1}}{1+{q'}^{2n-1} } \right)^4  \comma
\ee
where $q'$ is the nome conjugate to $q$. If we define $\lambda, u_1, u_2, u_3$ so that for $j=1, 2, 3$, 
\be q= \e^{-\pi \lambda} \sep z_j = \e^{-\pi u_j} \comma \ee
then, using (\ref{aprod}),
\be  q' \eq  \e^{-\pi /\lambda}  \sep u_1+u_2+u_3 = \lambda \period \ee

The system  is ferromagnetically ordered  if $z_1,z_2,z_3$ are real and positive and
\be  0 < k <1 \period \ee

 If  $ a_1 = a _2 = a_3 = q^{1/3}$, then we regain the isotropic case, with $q= p^6$.



 
 \subsection*{The bulk and surface free energies}
 

 Again we can use (11.7.14) of \cite{mybook} to obtain
 \be \label{kapb1}  \kappa_b \eq  \boldsymbol{\kappa}_b(a_1,q) \,   \boldsymbol{\kappa}_b(a_2,q) 
\,  \boldsymbol{\kappa}_b(a_3,q) \comma \ee
where
 \bd  \boldsymbol{\kappa}_b(a,q) \eq \prod_{k=1}^{\infty}  \left\{
 \frac{(1- q^{4k-2})^{2/3}}{(1- q^{4k-3})^{1/3}(1- q^{4k-1})^{1/3}} \, 
 \left[  \frac{(1- q^{4k - 2}/a)}{(1- q^{4k - 2} a) } \right]^{2k-1}
  \right. \ed
\be \label{kapb2}
\negspce  \left. \left[  \frac{(1-q^{4 k  + 1} a )(1-q^{4 k - 1} a )(1-q^{4 k}/a)}
{(1-q^{4 k -1} / a )(1-q^{4 k + 1} /a )(1-q^{4 k} a)} \right]^{2k} \right\} \comma \ee
or equivalently
\be \label{formkb}
\boldsymbol{\kappa}_b(a,q) \eq \left( \frac{4 q^{1/2}}{k} \right)^{1/12} \,
\left( \frac{G(a,q)}{H(a,q)} \right)^{1/2}  \ee
When $a_1=a_2=a_3 = q^{1/3}$ and $q=p^6$,  this is the same as (\ref{kpb}) above.


We could presumably derive at least the surface free energies by generalizing the methods 
of \cite{RJB2017}.We have not done so, but from our series expansions (up to 
order $p^{108}$) we conjecture  that
\be \label{kapsi}  \kappa_{s,i} =   \boldsymbol{\kappa}_{s}(a_i | a_{i+1},a_{i-1}|q) \comma \ee
where
\bd  \boldsymbol{\kappa}_{s} (a_1| a_2,a_3|q) \eq \prod_{k=1}^{\infty} \left(  \frac{(1-q^{2k- 1/2}/ a_1^{1/2})
(1-q^{4k-3}a_1) } {(1-q^{2k- 3/2} a_1^{1/2} )(1-q^{4 k -1}/a_1) } \right)^{2k-1} 
 \left( \frac{1-q^{2 k-1/2} a_1^{1/2} } {1-q^{2 k + 1/2}/a_1^{1/2}} \right)^{2k} \ed
\bd  \times  \left( \frac{(1-q^{4 k}/a_1)( 1-q^{4 k+2}/a_1)}
{(1-q^{4 k} a_1) ( 1-q^{4 k-2} a_1)} \right)^{k}   \left( \frac{(1-q^{2 k} a_2)( 1-q^{2 k} a_3)}{(1-q^{2 k} / a_2)
( 1-q^{2 k} / a_3)} \right)^{k/2}  \ed
\be \label{kaps2}
 \times \left( \frac{(1-q^{4 k -1}/a_2)(1-q^{4 k +1}/a_2)((1-q^{4 k -1}/a_3)
(1-q^{4 k +1}/a_3)}{
(1-q^{4 k + 1} a_2)(1-q^{4 k -1} a_2)(1-q^{4 k +1} a_3 )((1-q^{4 k -1} a_3)} \right)^k 
\comma \ee
or equivalently
\bd  \boldsymbol{\kappa}_{s} (a_1| a_2,a_3|q) \eq H(a_1^{1/2}/q^{1/2},q) \,
\left( \frac{H(a_1/q,q)H(a_2,q) H(a_2,q)}{G(a_1/q,q) H(a_1/q,q^2)^2 G(a_2,q) G(a_3,q) }\right)^{1/4}
\period \ed

 \subsection*{The corner  free energies}

For the $60^{\circ}$ corner free energy we can choose


\be \label{kapci1}
\kappa_{c,i} \eq  \boldsymbol{\kappa}_c(a_i,q)  \comma \ee
where
\be  \label{kapci2}
  \boldsymbol{\kappa}_{c}(a,q) =  \left(  \prod_{k=1}^{\infty}   \frac{  (1-q^{4 k-2} ) ^{1/6} (1-q^{2 k-1} ) ^{2k-1}}
{(1-q^{4 k-2})^{5k-5/2} \,  (1-q^{4 k})^{3k} } \right) \, \left( \prod_{m=1}^{\infty} \frac{ 1}{(1-q^{m-1/2}/a^{m-1/2})} \right) \ee
\bd 
\times \,  \prod_{k=1}^{\infty}   \prod_{m =1}^{\infty}  \frac{ (1-q^{2 k +m/2-1} /a^{m/2})^{4k-2} } 
{(1-q^{2 k+m-1/2 }/a^{m-1/2})^{4k} (1-q^{4k+m-2}/a^{m})^{10k-5}  (1-q^{4k+m} /a^{m})^{6 k } }
\period \ed

The $120^{\circ}$ corner free energy is the most cumbersome of all our formulae. With the above choice of 
$\kappa_{c,i}$ it is
\bd \kt_{c,i} \eq   \boldsymbol{\tilde{\kappa}}_c(a_i | a_{i+1}, a_{i-1}|q) \comma \ed

\noindent where
 \be \label{breakkappac}
 \boldsymbol{\tilde{\kappa}}_c(a_1 | a_{2}, a_{3}|q)  \eq P_0 \, P_1(a_1) P_2 (a_1) P_3(a_1) P_4(a_1 | a_2,a_3) P_5(a_1 | a_2,a_3) 
\ee
and,  defining 
\bd \epsilon_{m,j} \eq  1/2  \;  \; {\myrm{if} }  \; m = j \sep  \epsilon_{m,j}  \eq 1 \;  \; {\myrm{if } } \;  m > j \comma \ed 
\bd
 P_4(a_1 | a_2,a_3)   \eq  Q(a_1,a_2) Q(a_1,a_3)   \sep  P_5(a_1 | a_2,a_3)   \eq  R(a_1,a_2) R(a_1,a_3) \comma 
\ed

\noindent we have

\bd  \negspce \negspce  \negspce \negspce   \negspce \! \! \! \! P_0 \eq  \prod_{k=1}^{\infty}  (1-q^{4k-2})  \comma \ed

 \bd P_1(a)  \eq  \prod_{ k=1}^{\infty}  \prod_{ m=1}^{\infty}  \frac{(1- a^{m-1/2} q^{2 k +m-3/2})^{4k-2} (1- a^{m-1/2} q^{2 k+ m-1} )^2}{(1- a^{m-1/2} 
 q^{2 k+m-5/2})^{4k-4} (1- a^{m-1/2} q^{2 k+m-2} )^2} \ed
 
 \bd \negspce P_2(a)  \eq  \prod_{ k=1}^{\infty}  \prod_{ m=0}^{\infty}  \frac{(1- a^{m} q^{2 k+m-1/2})^{2 \epsilon_{m,0}}}{(1- a^{m} q^{2 k+m-3/2})^{2 \epsilon_{m,0}} } \ed

\bd P_3(a]  \eq  \prod_{ k=1}^{\infty}  \frac{ {(1-q^{2k}/a)}^{k/2}}{ {(1-q^{4k-1}/a)}^{k} {(1-q^{4k+1}/a)}^{k} }   \; \times \ed
\bd  \frac{ {(1-q^{4k-3})}^{8k-5}{(1-q^{4k-1})}^{8k-2}}{ {(1-q^{4k-2})}^{9k-7/2} {(1-q^{4k})}^{7k} } \;  \times \ed
 \bd \frac{ {(1-a q^{4k-2})}^{15k-21/2}   {(1-a q^{4k})}^{15k-5} }
{ {(1-a q^{4k-1})}^{17k-9}   {(1-a q^{4k+1})}^{13k} } \;  \times \ed
\be \label{functionP3} \prod_{ m=2}^{\infty}  \frac{ {(1-a^m q^{4 k+m -3}) }^{16k-11}  { (1-a^m q^{4 k+m -1}) }^{16k-5} }
{ {(1-a^m q^{4 k+m}) }^{14k}  {(1-a^m q^{4 k+m-2}) }^{18k-9} } \comma  \ee
\bd  \negspce \negspce  \negspce \negspce   \negspce \negspce  Q(a,b) \eq  \prod_{ k=1}^{\infty}
 \frac{ {(1-b^{1/2}  q^{2k-3/2} )}^{2k-1} }{ {(1-b^{1/2}  q^{2k-1/2} )}^{2k} } \times \ed
\be \prod_{m=1}^{\infty} \frac{
 {(1-b^{1/2} a^{m/2}  q^{2k+m/2-3/2} )}^{(4k-2) \epsilon_{m,1} }  {(1-b^{1/2} a^{m/2}  q^{2k+m/2-1} )}
}{{(1-b^{1/2}  a^{m/2} q^{2k+m/2-1/2} )}^{4 k  \, \epsilon_{m,1}}
 \; {(1-b^{1/2} a^{m/2}  q^{2k+m/2-2} )} } \comma  \ee

  \be  \label{prdr} R(a,b)  \eq    \prod_{ k=1}^{\infty} \frac{ {(1-b \, q^{4k-4} )}^{k-1} {(1-b \, q^{4k -2})}^{k}  
 {( 1- b \, a \, q^{4k-3})}^{3k-2}   }{{(1-b \, q^{4k -3})}^{2k-1}   {(1-b\, a \,  q^{4k-2})}^{4k-2}   }  \ee
 \bd
  \times    \; \frac{{( 1- b\,  a \,  q^{4k-1})}^{3k-1} }{ {( 1- b \, a  \, q^{4k})}^{2k} } 
  \prod_{ m=2}^{\infty}  \frac{{(1- b \, a^{m} q^{4 k+m-4})}^{4k-3} {(1- b \, a^{m} q^{4 k+m-2})}^{4k-1} }
 {{(1- b \, a^{m} q^{4 k+ m-3})}^{6k-3} {(1- b \, a^{m} q^{4 k+m-1})}^{2k} }
 \period   \ed
 
 As discussed above, the choice of these functions $  \boldsymbol{\kappa}_{c}(a_i) ,   
  \boldsymbol{\tilde{\kappa}}(a_i | a_{i+1}, a_{i-1}) $ is not unique, though the product
  $  \boldsymbol{\kappa}_{c}(a_i)   
  \boldsymbol{\tilde{\kappa}}(a_i | a_{i+1}, a_{i-1}) $ is. We are free to multiply $  \boldsymbol{\kappa}_{c}(a) $ by 
  $u(a_i | a_{i+1}, a_{i-1}) $, and divide
  $  \boldsymbol{\tilde{\kappa}}(a_i | a_{i+1}, a_{i-1}) $ by   $ u(a_i | a_{i+1}, a_{i-1}) $, 
  provided only that 
  \bd  u(a_1 | a_2, a_3) \, u(a_2 | a_3, a_1) \,   u(a_3 | a_1, a_2)   = 1 \period  \ed

Apart from the square lattice Ising model result and the conjectures of Vernier and Jacobsen\cite{VJ2012},
the author knows of no argument that 
corner  free energies should have simple product expansions in terms of elliptic variables
such as $ a_1,a_2, a_3, q$ . However, we note  that the second argument $b$ only enters the 
factors in the function $Q$ linearly in $b^{1/2}$, and $R$ linearly in $b$.
This is a significant simplification and encourages us to believe in the correctness 
of these conjectures.


\subsection*{Summation formulae for the anisotropic case}
\label{sumsection}

As in the previous section, we can convert these products to sums by taking the
logarithms, obtaining

\be \label{kapeffb}
\log  \, \boldsymbol{\kappa}_b(a,q) 
 \eq  \sum_{m=1}^{\infty}  \frac{ {\boldsymbol{F}}_b (a^m,q^m) }{m} 
\comma \ee
where
\be  \label{FB}
 {\boldsymbol{F}}_b(a,q)  =\frac{q-q^2}{3 (1+q)(1+q^2)}   +\frac{q (a-1/a)}{2 (1+q^2)} 
- \frac{q (a-1/a)}{2 (1+q)^2} \period \ee
The series (\ref{kapeffb}) is convergent provided $ q^2 < |a| < q^{-2}$.


Similarly,  
\be  \label{logks}
\log  \boldsymbol{\kappa}_s (a_1| a_2,a_3|q) 
 \eq  \sum_{m=1}^{\infty}  \frac{ {\boldsymbol{F}}_s (a_1^m|a_2^m,a_3^m|q^m) }{m}  \comma \ee
 where
\bd     {\boldsymbol{F}}_s (a_1|a_2,a_3|q)  \eq \frac{q^{1/2} (a_1^{1/2}-q/a_1^{1/2})}{(1+q)^2} +
\frac{a_1-q^2/a_1}{4 (1+q)^2} \; -
\frac{(1+q)^2 (a_1-q^2/a_1)}{4 (1+q^2)^2} + \ed
\be    \label{FS} \frac{q(a_2-1/a_2+a_3-1/a_3)}{4 (1+q)^2}
 -  \frac{q(a_2-1/a_2+a_3-1/a_3)}{4 (1+q^2)}  \comma \ee
 the series (\ref{logks}) being convergent if $q^3 < |a_1| < 1/q$ 
 and $q^2 < |a_2|, |a_3| < 1/q^2$.

Also, for  $|a| > q$, 
\be \label{logkc}
 \log  \boldsymbol{\kappa}_c(a,q)
 \eq  \sum_{m=1}^{\infty}  \frac{ {\boldsymbol{F}}_c (a^m,q^m) }{m} 
\comma \ee
where
\bd  {\boldsymbol{F}}_c(a,q) \eq -\frac{q^2}{6(1-q^4)}  + 
\frac{q^{1/2} (1+q^2) }{a^{1/2} (1- q/a )(1+q)^2)}  \; \; -   \ed
\be \label{defFc}
 \frac{q (1+q/a)}{  
 (1-q/a)(1+q)^2}  + \frac{q^2 (1+q/a) }{2  (1-q/a) (1+q^2)^2 } \period \ee

Finally, provided  $|a_1| < q^{-1}$ and $|a_2|,|a_3| < {\myrm{min}}[q^{-1}, q^{-1} |a_1|^{-1}]$, 
\be   \label{logkappact}
\log \boldsymbol{\tilde{\kappa}}_c(a_i | a_{i+1}, a_{i-1}|q) 
 \eq  \sum_{m=1}^{\infty}  \frac{ \Ft_c (a_1^m|a_2^m,a_3^m | q^m) }{m}  \comma \ee
where
   \bd \Ft_c(a_1 | a_2,a_3 | q) \eq -\frac{q^2}{3(1-q^4)}  + \frac{2 \,  q \,  a_1^{1/2} \,  
 (1-q^{1/2}+q)}{(1+q)^2 (1-q \,  a_1)}  +  \frac{q^{1/2} (1+q \, a_1)}{(1+q)(1-q \, a_1) }\; + \ed
 \bd   \frac{q^2 (1+q \, a_1)}{2  (1+q^2)^2 (1-q \,a_1)} - 
 \frac{2 q (1+q a_1)}{ (1+q)^2 (1-q\, a_1)} -\frac{q(1+q a_1)}{2 (1+q^2)(1-q a_1)} \; -\ed
 \bd \frac{q (1-q \, a_1)(q+a_1)}{2 a_1 (1+q)^2(1+q^2)}  -
  \frac{q^{1/2} (a_2^{1/2}+a_3^{1/2}) (1-a_1^{1/2})
   (1-q\, a_1^{1/2})}{(1+q)^2(1-q^{1/2} \, a_1^{1/2})}  \; +\ed
   \be \label{effct}
   \frac{q (a_2+a_3)(1+q+q^2)(1-a_1)(1-q^2 \, a_1)}{ (1+q)^2(1+q^2)^2 (1-q \, a_1)} \period \ee
   
   These formulae for ${\boldsymbol{F}}_b, {\boldsymbol{F}}_s, {\boldsymbol{F}}_c, 
   \Ft_c$ follow from the product  forms (\ref{kapb2}) - (\ref{prdr}). They all have the property 
  [using (\ref{aprod})] that they are negated by inverting their arguments. When 
   $a_1 = a_2 = a_3 = q^{1/3}$ they of course agree with the isotropic formulae
   (\ref{logk})  - (\ref{fkct}).

We are taking $0 < q <1$  throughout this paper, so the products in 
(\ref{kapb2}) - (\ref{prdr}) are always convergent. The sums in (\ref{FB}) - 
(\ref{effct}) are convergent only if the specified restrictions are satisfied.




\section{Inversion-type relations}
\setcounter{equation}{0}
The results of the  previous two sections should apply throughout the physical 
ferromagnetic regime, where $z_1,z_2,z_3$ are all real, positive, and less than one.
This is when $a_1, a_2, a_3$ are all real and 
\be p^4 <  a_j < p^{-2}  \; \; \; {\myrm{and} } \; \; \;  a_1 a_2 a_3 =1 \period \ee
However, our results for $  \boldsymbol{\kappa}_b(a) , \ldots , 
 \boldsymbol{\tilde{\kappa}}_c(a_1 | a_{2}, a_{3}) $ are meromorphic functions of 
 $a_1,a_2,a_3$ in the complex plane, so can immediately be extended to all
 complex values. It is these extensions (analytic continuations) that we consider here.

The edge matrices $U_{j,j+1}, V_j, W_{j,j+1}$  depend on $K_1, K_2, K_3$, or equivalently 
$z_1, z_2, z_ 3$,  respectively. {From} (\ref{defUVW}),
\bd [ {U_{j,j+1}(z_1) }]^{\, -1} = U_{j,j+1}(z_1^{-1}) \sep  V_j (z_2) ^{-1} = (1-z_2^2)^{-1} \,
 V_j(-z_2) \ed
\be  [ {W_{j,j+1}(z_3) }]^{\, -1} = W_{j,j+1}(z_3^{-1}) \period  \ee

If we write  the transfer matrix $T$ as $T(z_1,z_2,z_3)$, then from (\ref{defT}), its 
transposed inverse is
\be   {T(z_1,z_2,z_3)^{-1}}^{\dagger} \eq (1-z_2^2)^{-N} \, T(1/z_1,-z_2,1/z_3) \period  
\ee

If $\Lambda(z_1,z_2,z_3)$ is the maximum eigvalue of $T$ in the physical regime 
(when $z_1, z_2, z_3$ are real, positive and less than one), then this suggests that 
the analytic continuation of  $\Lambda(z_1,z_2,z_3)$  satisfies
\be \Lambda(z_1,z_2,z_3) \; \Lambda(1/z_1,-z_2,1/z_3) \eq (1-z_2^2)^{N}  \period \ee

Since 
\bd \Lambda(z_1,z_2,z_3) \eq \kappa_b^N \, {\kappa_{s,2}}^2 \comma \ed
this implies, using the $z_1,z_2 \rightarrow z_2,z_1$ symmetry, that 
\bd \kappa_b(z_1,z_2,z_3) \, \kappa_b(1/z_1, -z_2, 1/ z_3)  \eq 1-z_2 ^2 \comma \ed
\be \label{kapps}
 \kappa_{s,1}(z_1,z_2,z_3) \,  \kappa_{s,1}(-z_1, 1/z_2, 1/ z_3)  \eq 1 \period \ee
 
 Using (\ref{defzj}), we can verify that  
 \be z(1/ a,q)  \eq 1/z(a,q) \sep z(q^2/ a,q) \eq -z(a,q)  \period \ee
 Changing variables from $z_1, z_2 , z_3$ to $a_1,a_2, a_3$, 
 from (\ref{kapb2}), (\ref{kapsi}) and (\ref{kapps}), 
 \be \label{inv1}
 \boldsymbol{\kappa}_b(1/a,q)  \,  \boldsymbol{\kappa}_b(a,q) \eq \chi  \sep 
  \boldsymbol{\kappa}_b(q^2/a,q)  \,  \boldsymbol{\kappa}_b(a,q) \eq \frac{1-z(a,q)^2}{\chi^2}  \comma \ee
  \be  \label{inv2}
  \boldsymbol{\kappa}_{s}(a_1| a_2,a_3) \,  \boldsymbol{\kappa}_{s}(q^2/a_1| 1/a_2,1/a_3) \eq 
  1 \comma \ee
  where $\chi$ is a constant, independent of $a_1,a_2, a_3$, but possibly dependent on $p$.
  (\ref{inv1}) and (\ref{inv2}) are ``inversion relations''.\cite{Strog, Bax82,RJB2017} For the square lattice, we were able to obtain 
  two more relations (one for $\kappa_{s,i}$ and one for $\kappa_{c}$), but unfortunately
  that method breaks down here.\footnote{We cannot write $Z$ in the form (7.3) of 
 \cite{RJB2017},  the vector  $\xi$ being independent of $a_1,a_2,a_3$.}
 
 
 \subsection*{Observed  identities}
 
Now we use our results of section (\ref{sumsection}) to see if they do in fact satisfy 
(\ref{inv1}), (\ref{inv2}) and any similar relations. 

Define 
\be \eta =   \prod_{k=1}^{\infty} \frac{(1-q^{4 k-2})^2}{(1-q^{2 k-1}) } \sep \mu = 
\prod_{k=1}^{\infty} (1-q^{4 k-2}) \period \ee
Using the theory of elliptic functions, we can establish from  (\ref{defz}) that
\be 1- z(a,q) ^2 \eq  \prod_{k=1}^{\infty} \frac{(1-q^{4 k-2})^4 (1-q^{2 k-2} a)
 (1-q^{2 k}/a)}{(1-q^{2 k-1})^2 \,  (1-q^{4 k-3} a)^2 \, (1-q^{4 k-1}/a)^2} \comma \ee
 \be \log \eta \eq \sum_{m=1}^{\infty}  \left(\frac{ q^{m} }{m(1+q^{m} )} -\frac{q^{2 m}}{m(1+q^{2 m})} \right)
 \comma \ee
\bd \log \mu \eq -  \sum_{m=1}^{\infty} \, \frac{ q^{2m} }{m(1-q^{4m} )} 
 \comma \ed
and
 \be \log [1-z(a,q)^2 ]  = \sum_{m=1}^{\infty}   \left(2 -
 \frac{q^{m}}{a^m} -\frac{a^m}{q^{m}} \right )  \left(\frac{ q^{ m} }{m(1+q^{m} )} -\frac{q^{2 m}}{m(1+q^{2 m})} \right)
 \comma \ee

Now look for relations between the $ {\boldsymbol{F}}$ functions  defined above. We find three relations involving $ {\boldsymbol{F}}_b$ and $ {\boldsymbol{F}}_s$:
\be  {\boldsymbol{F}}_b(a,q)  + {\boldsymbol{F}}_b(1/a,q)  \eq  \frac{2}{3}   \left(\frac{ q }{1+q } -\frac{q^{2}}{1+q^{2}} \right) 
\comma \ee
\be  {\boldsymbol{F}}_b(a,q)  + {\boldsymbol{F}}_b(q^2/a,q)  \eq    \left( \frac{2}{3} -
 \frac{q}{a} -\frac{a}{q} \right )  \left(\frac{ q }{1+q } -\frac{q^{2}}{1+q^{2}} \right)
  \comma \ee

\be  {\boldsymbol{F}}_s (a_1|a_2,a_3 | q)  +  {\boldsymbol{F}}_s(q^2/a_1
| 1/a_2,1/a_3 | q )  \eq  0  \period \ee

 \noindent The corresponding sums in (\ref{FB}), (\ref{logks}) are convergent in the physical region $q < a,a_1,a_2,a_3 < 1$ (and beyond). Together with the relations of section (\ref{sumsection}), these relations  imply that
 \be  \label{ktinv1}
  \boldsymbol{\kappa}_b(a,q) \,   \boldsymbol{\kappa}_b(1/a,q) \eq \eta^{2/3} \sep
\  \boldsymbol{\kappa}_b(a,q) \,   \boldsymbol{\kappa}_b (q^2/a,q) \eq\frac{ 1-z(a)^2 }{ \eta^{4/3} } \comma \ee
 \be    \label{ktinv2}  \boldsymbol{\kappa}_{s}(a_1| a_2,a_3 |q)  \,  \boldsymbol{\kappa}_{s}(q^2/a_1| 1/a_2,1/a_3 | q)  \eq 1 \period  \ee
 We see that these  are indeed the expected inversion relations
(\ref{inv1}), (\ref{inv2}),  with $\chi =  \eta^{2/3}$.

 For the corner free energies, we also note from (\ref{defFc}) and (\ref{effct}) 
 the following two formal identities
  \be \label{tricky1}
    {\boldsymbol{F}} _c (a,q)  +  {\boldsymbol{F}} _c (q^2/a,q) \eq - 
    \frac{q^2}{3 (1-q^4)} \comma \ee

\bd {\boldsymbol{\tilde{F}}} _c (a_1|a_2,a_3 | q)  +  {\boldsymbol{\tilde{F}}}_c  
\left( \frac{1}{q^2 a_1} \left| \frac{q^2}{a2},\frac{q^2}{a3}  \right| q \right) 
\eq  -\frac{2 q^{2}}{3 (1-q^{4})}  \;  + \ed
\be  \label{tricky2} \frac{(1-q) (1-q a_1)^2  }{2 a_1 (1+q) (1+q^2)}  \period \ee

\noindent However, there is no domain within which the corresponding series (\ref{logkc}), 
(\ref{logkappact}) are convergent for both terms in (\ref{tricky1}) and (\ref{tricky2}). 
(The inversion (overlap) points are $a=q$ and $a_1=q^{-1},   \, a_2 = a_3 = q$, which are 
at the boundary of the convergence of (\ref{logkc}), (\ref{logkappact}).)


If one does naively substitute these formulae into the series, one obtains
  \be  \label{ktinv33} \boldsymbol{\kappa}_{c}(a,q)  \, \boldsymbol{\kappa}_{c}(q^2/a,q) 
   \eq  \mu^{1/3}  \comma \ee
 \ba   \label{ktinv4}  \negspce \boldsymbol{\tilde{\kappa}}_c(a_1 | a_{2}, a_{3} | q) 
  \!  \!  \!  & &  \!  \!  \!  \!  \!  \!   \boldsymbol{\tilde{\kappa}}_c
 \left( \frac{1}{q^2 a_1} \left| \frac{q^2}{a2},\frac{q^2}{a3}  \right| q \right)   = 
\frac{ \mu^{2/3} \,  z(a_1,q) }{[z(a_1,q)^2- 1]^{1/2}}   \comma \ea
for $1<  |a_1| < q^{-1} $, but these equations  appear to be meaningless.
 $ \boldsymbol{\kappa}_{c}$  has an essential singularities at $a = q$, so cannot be 
 analytically continued to smaller values.
Similarly,  $ \boldsymbol{\tilde{\kappa}}_c$ cannot be continued to $a> q^{-1}$.

To use the  inversion relation technique given in section 7 of \cite{RJB2017}, we 
need two distinct inversion-type relations to define a function, together with its being 
analytically continuable from one side of each inversion point to the other. 
Hence (\ref{ktinv1}) is sufficient to determine
$ \boldsymbol{\kappa}_b(a,q) $, but (\ref{ktinv2}) is not sufficient to  determine
$\boldsymbol{\kappa}_{s}(a_1| a_2,a_3 |q) $.

Nor  would the problematic equations (\ref{ktinv33}), (\ref{ktinv4})  be sufficient to
 determine $\label{ktinv3} \boldsymbol{\kappa}_{c}(a,q)$
or $ \boldsymbol{\tilde{\kappa}}_c(a_1 | a_{2}, a_{3}|q)$.



\section{Behaviour near criticality}
\setcounter{equation}{0}
\setcounter{equation}{0}
We define $\lambda, u_j, q'$ (for $j= 1, 2, 3$)  by
\be \label{defqp}
q= \e^{-\pi \lambda} \sep a_j = \e^{-\pi u_j} \sep
 q' = \e^{-\pi/\lambda} \comma \ee
 and note from (\ref{restra}) that
 \be 0 \leq  u_j < \lambda \ee
 for $j=1, 2, 3$.

In  eqns. (\ref{defGwq}) - (\ref{logH}) we set
\be w \eq \e^{-\pi u }\period \ee
The critical case is obtained  by taking the limit  
\be \lambda, u_j  \rightarrow 0^{+}  \ee
keeping the ratios $u_j/\lambda$ fixed. (For the isotropic case  they are  1/3.).
Then $q, k \rightarrow 1$ and the system becomes critical.

In this limit the above products and sums all converge more and more
slowly. However, fortunately we can transform them to forms that 
converge quickly by  using either the conjugate modulus
identities of elliptic functions, or more generally by using the 
Poisson transform\cite[eq. 15.8,1]{mybook}.

\subsection*{Poisson transform}
If $g(x)$ is analytic on the real axis and its Fourier
transform 
\be \label{four}
 \widehat{g}(y) \eq \int_{_\infty}^{\infty} \e^{\i x y } \, g(x) \, dx \ee
exists for all real $y$, then for any positive $\delta$,
\be \label{PT}
\sum_{n= -\infty}^{\infty} g(n\delta) \eq \delta^{-1} \sum_{n=-\infty}^{\infty} 
\widehat{g}(2 \pi n/\delta)  \period \ee
All the functions $g(x)$ that we shall deal with are even
functions, with $g(-x)=g(x)$, so $\widehat{g}(-y)=\widehat{g}(y)$ and 
(\ref{PT})  can be written
\be \label{PTf}
 \sum_{n=1}^{\infty} g(n \delta) \eq - g(0)/2 +  \widehat{g}(0)/(2\delta ) +
 \delta^{-1} \sum_{n=1}^{\infty} 
\widehat{g}(2 \pi n/\delta)  \period \ee

\subsection*{The Boltzmann weights $z_1, z_2, z_3$ }
The function $w^{1/2} G(w,q)$, where $G(w,q)$ is defined by
(\ref{defGwq}), is the ratio of two elliptic theta functions. Either using the
conjugate function identities of (15.7.1) - (15.7.3) of \cite{mybook}, or using
the Poisson transform above, we find that (for all $w,q$)
\bd \log G(w,q) \eq \pi u/2 +\log \tan\left[ \frac{\pi (\lambda-u)}{ 4\, \lambda} \right] - \,  4
 \sum_{m \, {\myrm{odd}}} \frac{(-1)^{(m-1)/2} \, q' \m \,\sin[ \pi m u/(2 \lambda) ] }
{m (1- q' \md) } \ed
where  $q'$ is defined by  (\ref{defqp}). The sum is over all odd  positive values of $m$, i.e.
$m= 1,3 , 5 , \ldots$  Hence from (\ref{defz2}),  for $j=1, 2, 3$,
\be \label{formzj}
\log \, z_j \eq \log \tan\left[ \frac{\pi (\lambda-u_j)}{ 4\, \lambda} \right] - \,  4
 \sum_{m \, {\myrm{odd}}} \frac{(-1)^{(m-1)/2} \,  q'  \m \,\sin[ \pi m u_j/(2 \lambda) ] }
{m (1-q'  \md) } \period \ee
We see that in the critical limit, when $q' \rightarrow 0 $, 
\be z_j \eq \tan\left[ \frac{\pi (1-u_j/\lambda)}{ 4} \right]  \times [ 1-
4q' \sin(\pi u_j/2 \lambda) + {\myrm{O}} (q'^2) ] \period \ee
So $q'$ is proportional to the deviation  from criticality $T_c-T$. We shall regard
$q'$ as that deviation, i.e. we take
\be T_c - T \eq  q' \period \ee

\subsection*{The function $H(w,q) $}

We shall also need the the critical behaviour of the function
$H(w,q)$ defined by (\ref{logH}), i.e.
\be 
\log H(w,q) \eq - \frac{\pi \lambda}{4} \, \sum_{n=1}^{\infty} g(n \pi \lambda/2) \comma \ee
where
\be g(x) \eq \frac{\sinh(2 ux/\lambda) }{x \cosh^2 (x) } \period \ee
Defining $g(0) = \lim_{\, x \rightarrow \infty} g(x) = 2 u/\lambda$, this $g(x)$ is analytic on the real 
axis and, provided $|u| < \lambda$, we can use the Poisson transform above.

In (\ref{four}), when $y >0 $ we can close the contour of integration round the upper half x-plane. The only 
singularities within the contour are double poles at
\bd x =  \pi m \, \i /2 \comma \ed
where $m$ is an odd integer. The associated residue of $\e^{\i xy} g(x)$ is
\bd R_m(y)  \eq \frac{- 2 \, \i \,  \e^{-\pi m y/2} \left[ 2 S_m - 2 \pi m u \,  C_m/\lambda +\pi m y S_m \right] }{\pi^2 m^2}
\comma \ed 
where \bd S_m = \sin( \pi m u/\lambda) \sep C_m = \cos( \pi m u/\lambda) \comma \ed
so for $y > 0 $,
\be \widehat{g}(y) \eq 2 \pi \i   \sum_{\, m \;  \myrm{odd} } R_m(y) \period \ee
Substituting into (\ref{PTf}) and performing the summation over $n$, we obtain
\ba \label{Hwq}
&& \log H(w,q) =  \pi u/4 -\widehat{g}(0)/4-\half \sum_{n=1}^{\infty} \widehat{g}(4n/\lambda) \nonumber  \\
\negspce \negspce & \negspce = & \! \! \! \! \! \! \pi u/4 - \! \widehat{g}(0)/4-   
 \! \! \! \sum_{\, m \; {\myrm{odd}}} \left[ \frac{4 \,  q' \mdb (S_m - \pi m u C_m/\lambda)}{\pi m^2 (1- q'\mdb)}  + 
 \frac{ 8  \, q'\mdb S_m}{\lambda m (1- q'\mdb)\rs} \right]  \period \ea

\subsection*{$ \boldsymbol{\kappa}_b$ and $ \boldsymbol{\kappa}_{s}$ }

{From} (15.1.4b) of \cite{mybook},
\be k \eq  \prod_{m \, {\myrm{odd}}} \, \left( \frac{1- q'\md}{1+q'\md} \right)^4 \ee
and from the equation before (\ref{formzj}),
\be G(w,q) \eq  \e^{\pi u/2} \, f_1(u/\lambda,q') \comma \ee
where we shall take $u/\lambda$ to be fixed and the function $f_1$ is analytic in $q'$ at 
$q'=0$. Also, from (\ref{Hwq}),
\be \label{hsing}
H(w,q) \eq  \e^{\pi u/4} \, g_2(u/\lambda,q'\rs) \left[1- \frac{8 q'\rs
 \sin(\pi u/\lambda) }{\lambda} + {\myrm{O}}( q'\rsq /\lambda)  \right]\comma \ee
 where this function $g_2(u/\lambda,q'\rs) $ is analytic in $q'\rs$ at $q'=0$.

Substituting these formulae into (\ref{formkb}), we obtain
\be \label{kba}
\boldsymbol{\kappa}_b(\e^{-\pi u}, \e^{-\pi \lambda} ) \eq 2^{1/6} \,  \e^{-\pi (\lambda-3u)/24} \, 
 \left[1+\frac{4 q'\rs
 \sin(\pi u/\lambda) }{\lambda} + {\myrm{O}}( q'\rsq /\lambda)  \right] \,g_3(u/\lambda,q') 
 \comma \ee
 where $g_3(u/\lambda,q') $ is non-zero and analytic  in $q' $ at $q'=0$.

Now from (\ref {kapb1}),
\be  \label{kbb}
\kappa_b \eq \prod_{j=1}^3  \boldsymbol{\kappa}_b(\e^{-\pi u_j})  \comma \ee
and from (\ref{aprod}), 
\be  \label{sumuj}
u_1+u_2+u_3 = \lambda  \period \ee
It follows that the contributions of the factor $ \e^{-\pi (\lambda-3u)/24} $ in (\ref{kba})
cancel out of  (\ref{kbb}), leaving the dominant singular contribution to the bulk free energy  
given by
\be - \beta \, (f_b)_{\myrm{sing}}  \eq -  \frac{4 q'\rs
 \, \log q' }{\pi}  \, \sum_{j=1}^3 \sin(\pi u_j/\lambda ) \period \ee
 
 The surface partition function per site $ \boldsymbol{\kappa}_{s}$ is conveniently 
 given by the equation after (\ref{kaps2}). Substituting the above forms of $G(w,q)$ 
 and $H(w,q)$, we find, again using (\ref{sumuj}),
 that the leading factors $\e^{\pi u_j/2},  \e^{\pi u_j/4}$ cancel one another out.
 
 The other factors are analytic (Taylor expandable) in $q'$, except for the singular parts of 
 the $H$ functions, given by (\ref{hsing}). The dominant one of these is the factor coming 
 from $H(a_1 q,q^2)$, which is (replacing $\lambda,q'$ in (\ref{hsing}) by 
 $2 \lambda, {q'}^{1/2}$)
 \bd 1- \frac{4 q' \sin[\pi (1+u_1/ \lambda)/2]}{\lambda} + {\myrm{O}}(q'\rs/\lambda) \ed
 and gives the dominant  singular contribution to the surface free energy:
 \be  - \beta \, (f_{s,i})_{\myrm{sing}}  \eq -  \frac{ q'
 \, \log q' }{\pi}  \, \sin[\pi (1+u_1/ \lambda)/2]  \period \ee

 \subsection*{$ \boldsymbol{\kappa}_c$ and $\boldsymbol{\tilde{\kappa}}_c$ }
{From} (\ref{logkc}) and  (\ref{defFc}), 
\be   \log \boldsymbol{\kappa}_{c}(\e^{-\pi u }, \e^{-\pi \lambda }) \eq \sum_{n=1}^{\infty} 
\tilde{g}(n) \comma \ee
where  
\be \tilde{g}(n) \eq {\boldsymbol{F}}_c(\e^{-\pi u n }, \e^{-\pi \lambda n}) /n \period \ee
This function $\tilde{g}(n)$ is even, but it is {\em not} analytic at $n = 0$. Rather,
 it has the expansion about $n=0$:\footnote{Note that because of (\ref{sumuj}), the constant term
 $ \pi (3 u-\lambda)/72$ cancels out of  the product
 $\kappa_{c,1} \, \kappa_{c,2}  \, \kappa_{c,3} $.}
 \be \label{ft1}
  \tilde{g}(n) \eq \frac{5 \lambda+u}{24 \pi \lambda (\lambda-u) n^2} + 
 \frac{\pi (3 u-\lambda)}{72} + {\myrm{O}}(n^2) \comma \ee
 so we cannot use the Poisson transform  of 
 (\ref{four}) - (\ref{PTf}).
 
 This difficulty is easily solved by defining
 \be  \tilde{g}(n) \eq \frac{5 \lambda+u)}{24 \pi \lambda (\lambda-u) n^2} +
   g(n)  \period \ee
 Using the formula
 \be \sum_{n=1}^{\infty} \frac{1}{n^2} \eq \frac{\pi^2}{6}  \ee
 \cite[48.2]{Dwight}, we see that
 \ba \label{lnkapc}
 \log \boldsymbol{\kappa}_{c}(\e^{-\pi u }, \e^{-\pi \lambda }) = -\beta f_c & = &
 \frac{\pi (5 \lambda+u)}{144 \lambda (\lambda-u) } +
 \sum_{n=1}^{\infty} g(n) \nonumber \\
  & = &
- \frac{(5 +u/\lambda) \log q' }{144 (1-u/\lambda) } +
 \sum_{n=1}^{\infty} g(n) \period \ea
 Taking $\delta =1 $, the last term on the RHS is again given by (\ref{PTf}), 
 with $\widehat{g}(0)$ a function only of $u/\lambda$.
 As $\lambda, u \rightarrow 0$, $g(0) \rightarrow 0$, $\widehat{g}(0)$ is a constant,
 and the sum of the RHS of  (\ref{PTf}) tends exponentially to zero.
 The dominant singularity is therefore given by the first term on the
 RHS of (\ref{lnkapc}) and is proportional to $q'$, i.e.to $T_c-T$.
 
 The calculation for $ \log \boldsymbol{\tilde{\kappa}}_c $ 
 follows closely that for  $  \log \boldsymbol{\kappa}_{c}$.
 {From} (\ref{logkappact}) and (\ref{effct}),
 \be \log \boldsymbol{\tilde{\kappa}}_c(\e^{-\pi u_1} | 
 \e^{-\pi u_2}, \e^{-\pi u_3}| e^{-\pi \lambda}) \eq
 \sum_{n=1}^{\infty} 
\tilde{g}(n) \comma \ee
where $\tilde{g}(n)$ is the RHS of (\ref{effct}), with
$a_j = \e^{-\pi u_j n}$ and $q = \e^{- \pi \lambda n}$, divided by $n$.
Again, using (\ref{sumuj}),  $\tilde{g}(n)$ is an even function, but has a double pole at 
$n=0$:
 \ba \label{ft2}
 \tilde{g}(n) & = & \frac{2 \lambda-u_1}{12 \pi \lambda (\lambda+u_1) n^2} + 
 \frac{\pi (\lambda-3 u_1)}{72} + {\myrm{O}}(n^2) \comma \nonumber \\
& = & \frac{2 \lambda-u_1}{12 \pi \lambda (\lambda+u_1) n^2} + 
g(n) \comma \ea
so 
 \be \label{lnkapct}
 \log \boldsymbol{\tilde{\kappa}}_c \eq -\beta \tilde{f}_c \eq - \frac{(2 -u_1/\lambda) \log q'}
 {72  (1+u_1/\lambda) } +  \sum_{n=1}^{\infty} g(n) \ee

 and the dominant singularity is given by the first term on the RHS.
  
 The bulk, surface and corner free energies therefore have singularities
 proportional to $(T_c -T)^2 \log (T_c-T),   (T_c -T) \log (T_c-T)$,  and 
 $\log (T_c-T)$, respectively, corresponding to the crititical exponents being
 $2, 1,  0$.

We observe the modularity property that the $g(0)$  terms in  (\ref{PTf}) 
cancel one another out of 
{\em all} the  allowed products of the $\kappa$'s
in (\ref{kapsums}) .

 \subsection*{Predictions of conformal invariance }
 
 For the isotropic triangular lattice, $u = u_1 = \lambda/3$ and 
 (\ref{lnkapc}), (\ref{lnkapct}) give
 \be  \label{ci1}
 - \beta f_c \eq   \frac{1}{18} \; \log q' \sep - \beta \tilde{f}_c \eq  
  \frac{5}{288} \; \log q' \period \ee
  We see in the next section that we can obtain the square lattice from the triangular
  by setting $K_3 =0$. This turns off the $K_3$ interactions. This square lattice model is 
  isotropic if $K_1 = K_2$, i.e. if 
  $a_1 =a_2 =q^{1/2},   \, a_3=1$ and hence $u_1= u_2= \lambda/2,\,  u_3 = 0$.
  The corners are now of type 3, so we should replace $u$ and $u_1$ in 
   (\ref{lnkapc}), (\ref{lnkapct}) by $u_3 =0$. Taking the arithmetic mean of those
   equations to remove the ambiguity discussed  at (\ref{transfm}), we obtain
   \be  \label{ci2}
   - \beta f_{c,sq} \eq \frac{1}{2}  \left( \frac{5}{144} +\frac{1}{36} \right) \log q' 
  \eq \frac{1}{32} \log q'  \period \ee
  
  These $f_c, \tilde{f}_c,   f_{c,sq}$ are the free energies of the corners of the isotropic
  $60^{\circ}$ triangular,   $120^{\circ}$  triangular,  and  $90^{\circ}$ square lattices, 
  respectively. In 1988 Cardy and Peschel\cite[eqn. 4.4]{CardyPeschel1988} predicted that 
 at the critical temperature  the divergent logarithmic term in the corner free energy of 
 any planar isotropic model of size $L$  would be
  \be \Delta F \eq \frac{c }{24 } \left( \frac{\gamma}{\pi} -\frac{\pi}{\gamma} \right) \, \log L  
  \period \ee
  where $\gamma$ is the internal angle of the corner and  $c$ is the conformal anomaly 
  number. For the planar spin 1/2 XY-model and the Ising model,  $c= 1/2$.\cite{Blote1986} 
  
  Taking $\gamma= \pi/3,  2\pi/3$ and $\pi/2$ in this formula, we obtain agreement with 
  the three equations in  (\ref{ci1}), (\ref{ci2}), provided we replace $-\log q'$ by 
  $\log L$.\footnote{See equations 93 and 94 of Vernier and Jacobsen.\cite{VJ2012} }




\section{Comparison with the square lattice}

\setcounter{equation}{0}
   
   The square lattice may be regarded as a special case of the triangular lattice, 
 when one of the three interactions is turned off. if we take
 \be a_3 = 1 \comma \ee
 then from (\ref{defGwq}) and (\ref{defz2}), (\ref{defzj})
 \bd z_3 =1 \sep K_3 = 0 \period \ed
 This is equivalent to removing the SE-NW lines in the first 
diagram in Fig. \ref{examples}, so it just becomes a tilted  version of the
square lattice we discussed in \cite{RJB2017}. The $K_1, K_2$ of this paper are the
$H' , H$  of  \cite{RJB2017}, and from eqns. (2.7), (2.23), (6.8), (6.26) therein,
\bd \e^{-2H'} \eq \frac{q^{1/2}}{w} \, G(q/w^2,q) \sep 
\e^{-2H} \eq  w \, G(w^2,q)  \comma \ed
where $G(a,q)$ is defined by (\ref{defGwq}) above.

Comparing these equations with (\ref{defz2}) above, 
the $w$ of  \cite{RJB2017} is given in terms of our $a_1,a_2$ by
\be w= a_2^{1/2} = (q/a_1)^{1/2}  \ee
and we can compare our results for  $a_1 = q/w^2,   \, a_2=w^2,   \, a_3=1$ with 
eqn (6.36) of \cite{RJB2017} .

{From} (\ref{FB}) above,
\be F_b(q/w^2,q) + F_b(w^2,q) +F_b(1,q) \eq 
\frac{q (1-q) (w-q/w) (w^{-1}-w)}{(1+q)^2 (1+q^2)} \comma \ee
so from (\ref{kapb1} ) and (\ref{kapeffb}),
\be -\beta f_b \eq  \log \kappa_b \eq \sum_{m=1}^{\infty} 
\frac{q^m (1-q^m) (w^m-q^m/w^m) (w^{-m}-w^m)}{m(1+q^m)^2 (1+q^{2m})} \period \ee

Allowing for the fact that we are working with the partition function $\widehat{Z}$, whereas
 \cite{RJB2017} works with the full partition function $Z$, related by (\ref{ZtZ}), this is the 
same as the first of the equations  (6.36) of \cite{RJB2017}.

{From}  (\ref{FS}), 
\be  {\boldsymbol{F}}_s (q/w^2| w^2,1|q)  \eq  \frac{q (w^{-1}-w)}{(1+q)^2}
-  \frac{q^2 (w^{-2}-w^2)}{2(1+q^2)^2} \comma \ee
so 
\ba \label{odd&even} \! \! \log \kappa_{s,1}     \! \!  \! \! &  =  &   \! \!  \! \!  
\log  \boldsymbol{\kappa}_s (a_1| a_2,a_3|q) = 
\! \sum_{m=1}^{\infty} \! \frac{q^m (w^{-m} \minus w^m)}{m(1+q^m)^2}
-  \! \sum_{m=1}^{\infty}  \! \frac{q^{2m} (w^{-2m}\minus w^{2m})}{2m(1+q^{2m})^2}
\nonumber \\
 & \negspce = \negspce &  \sum_{m=1}^{\infty}  \frac{q^m (w^{-m}-w^m)}{m(1+q^m)^2}
\; -\sum_{m \; {\myrm{even}}}  \frac{q^m (w^{-m}-w^m)}{m(1+q^m)^2} \nonumber \\
 & \negspce = \negspce & \sum_{m \; {\myrm{odd}}} \frac{q^m (w^{-m}-w^m)}{m(1+q^m)^2}
 \period  \ea

{From} the first diagram of 
Fig.\ref{examples},  we see that $n_{c,1}$ is $ 2N$, so
comparing this with (1.1) of \cite{RJB2017} ,
our  $\log \kappa_{s,1}$ should be one-half of the $H - \beta f'_s$ therein.
{From} the third equation of  (6.36), we see that this is indeed so.

The equivalence of our result for $\log \kappa_{s,2}$ with the 
$H'-\beta f_s$ of \cite{RJB2017}  follows immediately by interchanging
$a_1$ with $a_2$ and $w$ with $q^{1/2}/w$.

Finally, we consider the corner free energies in the square lattice case.
From the first diagram in Fig. \ref{examples},
all four corners are of type 3, so to compare our results with those of
\cite{RJB2017}  we must now take  ${a_1, a_2, a_3 } = {q/w^2,w^2,1}$.
Defining
\be \boldsymbol{F}_{c,sq}(q,w)  \eq  {\boldsymbol{F}}_c(a_3,q) +\Ft_c(a_3 | a_1,a_2 | q) 
 \comma \ee
then 
\be \boldsymbol{F}_{c,sq}(q,w)   \eq{ \cal{F}}(q) -{\cal{F}}(q^2)/2 \comma \ee
where
\be  {\cal{F}}(q) = \frac{2 q^{1/2} (1+q^2)}{(1+q)(1-q^2)} -q/(1-q^2) \period \ee
We note that $ \boldsymbol{F}_{c,sq}(q,w) $ is independent of $w$.
 {From} (\ref{logkc}) and (\ref{logkappact}), proceeding similarly to (\ref{odd&even} ),
\ba  \label{fcsq}
\; & \! \! \! \! & \log [\kappa_c(a_3,q) \,  \boldsymbol{\tilde{\kappa}}_c(a_3| a_{1}, a_{2}|q) ]
 =   \sum_{m=1}^{\infty}
\frac{\boldsymbol{F}_{c,sq}(q^m,w^m) }{m} \nonumber \\
& = & 
\sum_{m=1}^{\infty} \frac{{\cal{F}}(q^m)}{m} - \sum_{m=1}^{\infty} \frac{{\cal{F}}(q^{2m})}{2m}
\eq \sum_{m\, {\myrm{odd}}}
\frac{{\cal{F}}(q^m)  }{m}  \nonumber \\
& = & \frac{\log k'}{8} + 2 \sum_{ m\, {\myrm{odd}}} \, \frac{q^{m/2} (1+q^{2m})}
{m (1+q^m)(1-q^{2m})} \comma \ea
where we have used the relation
\bd \log k' \eq -8 \sum_{m \; {\myrm{odd}}} \, \frac{q^m}{m (1-q^{2m})} \period \ed

Allowing for the fact that (\ref{fcsq}) is the free energy for two corners, of $60^{\circ}$ and 
$120^{\circ}$, whereas  the RHS of (6.36) of \cite{RJB2017} is the sum of all the four 
corners  in Fig. \ref{examples} and includes the logarithm of the factor 2 in \ref{ZtZ}, 
(\ref{fcsq}) agrees with (6.36) of \cite{RJB2017}.




\section{Summary}

\setcounter{equation}{0}

We have used  Kaufman's spinor method to simplify the exact calculation of the
partition function $Z$ of the ferromagnetically ordered  Ising model for an 
arbitrary convex polygon drawn on
the triangular lattice. From this we we have been able to calculate series 
expansions (to 108 terms) of the bulk, surface and the two corner free energies.
We observe repeat patterns of period 24, enabling us to extrapolate and thereby 
conjecture the exact results. Our result for the bulk free energy agrees with the 
known result of Houtappel and others.\cite{Hout1950}-\cite{Stephenson1964}

We first consider the low-temperature isotropic case, when $T< T_c$  and all edges 
of the lattice have interaction coefficient $K$ and Boltzmann 
weight $z= \e^{-2K}$. We conjecture the surface and corner free
energies by extrapolation of the series. Our conjectures agree with that for the $60^{\circ}$
corners of Vernier and Jacobsen.\cite{VJ2012}

We then  consider the more general case when the three types of edges have
different interaction coefficients  $K_1, \, K_2,  \, K_3$ and again $T < T_c$.
The results are then more complicated (particularly so for the $120^{\circ}$
corner free energy), but we do see sufficient patterns in the series to confidently
predict the the free energies. They agree with those for the isotropic case.

In all the cases we have 
studied, we find the general formula (\ref{defkappas1}) holds, with
$\kappa_b, \kappa_{s,i}, \kappa_{c,i}, \kt_{c,i}$ given by (\ref{kapb1}), (\ref{kapb2}), 
(\ref{kaps2})  and (\ref{kapci1}) - (\ref{prdr}). 

As a further check on our conjectures, we find that if we take the third interaction 
coefficient $K_3$ to be zero, and hence $a_3 = 1, u_3=0$,
we do indeed regain the exactly known square lattice results of \cite{RJB2017}.

We obtain the critical behaviour of the bulk, surface and
the two corner free energies, and find logarithmic singularities corresponding to
the exponents $\alpha$ being 2, 1, 0, 0, respectively, in agreement
with the square lattice results.   The results for the corner free energies on the isotropic
triangular and square lattices agree with 
 the predictions of conformal invariance\cite[eqn. 4.4]{CardyPeschel1988} 
and later numerical results.\cite{Wu2013,Wu2015}




\section{Acknowledgement}

\setcounter{equation}{0}
The author is grateful to Jesper Jacobsen for helpful comments on the relation of these
results to Cardy's predictions from conformal invariance, and to Jacques Perk for 
pointing out ref.\cite{Wu2013}

 \end{document}